\documentclass[aip,amsmath,amssymb,reprint]{revtex4-1}

\usepackage[version=3]{mhchem} % Formula subscripts using \ce{}
\usepackage[T1]{fontenc} 
\usepackage[utf8x]{inputenc}
\usepackage{rotating}
\usepackage{caption2}
\usepackage{bm}        % \boldsymbol (bold math)
\usepackage{amssymb}   % \mathbb     (blackboard bold)
\usepackage{amsmath}
\usepackage{threeparttable}
\usepackage{cancel}
\usepackage{tablefootnote}
\begin{document}
\author{Paweł Tecmer}
\email{ptecmer@fizyka.umk.pl}
\author{Marta Gałyńska}
\author{Lena Szczuczko}
%\altaffiliation{Current address: Some other place, Othert\"own,Germany}
\author{Katharina Boguslawski}
%\altaffiliation{A shared footnote}
\email{k.boguslawski@fizyka.umk.pl}
%\fax{+123 (0)123 4445557}
\affiliation
{Institute of Physics, Faculty of Physics, Astronomy, and Informatics, Nicolaus Copernicus University in Toru\'{n}, Grudzi{a}dzka 5, 87-100 Toru\'{n}, Poland}
%\alsoaffiliation[Second    University]
%{Department of Chemistry, Second University, Nearby Town}
%\author{Susanne K. Laborator}
%\email{s.k.laborator@bigpharma.co}
%\affiliation[BigPharma]
%{Lead Discovery, BigPharma, Big Town, USA}

%%%%%%%%%%%%%%%%%%%%%%%%%%%%%%%%%%%%%%%%%%%%%%%%%%%%%%%%%%%%%%%%%%%%%
%% The document title should be given as usual. Some journals require
%% a running title from the author: this should be supplied as an
%% optional argument to \title.
%%%%%%%%%%%%%%%%%%%%%%%%%%%%%%%%%%%%%%%%%%%%%%%%%%%%%%%%%%%%%%%%%%%%%
\title
  {Geminal-based strategies for modeling large building blocks of organic electronic materials}

\begin{abstract}
  We elaborate on unconventional electronic structure methods based on geminals and their potential to advance 
  the rapidly developing field of organic photovoltaics (OPV).
  Specifically, we focus on the computational advantages of geminal-based methods over standard approaches and identify the critical aspects of OPV development. Examples are reliable and efficient computations of orbital energies, electronic spectra, and van-der-Waals interactions. 
  Geminal-based models can also be combined with quantum embedding techniques and a quantum information analysis of orbital interactions to gain a fundamental understanding of the electronic structures and properties of realistic OPV building blocks.
  Furthermore, other organic components present in, for instance, dye-sensitized solar cells (DSSC) represent another promising scope of application.
  Finally, we provide numerical examples predicting the properties of a small building block of OPV components and two carbazole-based dyes proposed as possible DSSC sensitizers.
  %Finally, we discuss some numerical calculations, applying geminal-based methods to predict properties of some small building blocks of OPV components like the naphthalene diimide model system, and two carbazole-based dyes proposed as possible sensitizers for DSSC.
  %(including benzodithiophene and dithienopyrrole bridges). 
\end{abstract}

%%%%%%%%%%%%%%%%%%%%%%%%%%%%%%%%%%%%%%%%%%%%%%%%%%%%%%%%%%%%%%%%%%%%%
%% Start the main part of the manuscript here.
%%%%%%%%%%%%%%%%%%%%%%%%%%%%%%%%%%%%%%%%%%%%%%%%%%%%%%%%%%%%%%%%%%%%%

\maketitle
%%%%%%%%%%%%%%%%%%%%%%%%%%%%%%%%%%%%%%%%%%%%%%%%%%%%%%%%%%%%%%%%%%%%%
%%% Introduction
%%%%%%%%%%%%%%%%%%%%%%%%%%%%%%%%%%%%%%%%%%%%%%%%%%%%%%%%%%%%%%%%%%%%% 
Organic photovoltaic (OPV) devices and dye-sensitized solar cells (DSSC) represent an up-and-coming technology. 
For instance, their ever-improving cell efficiency (over 19\%)~\cite{opv-19-percent} and performance lifetime, combined with low environmental impact and potential roll-to-roll manufacturing, make OPVs competitive with conventional silicon-based technologies. 
An additional advantage of OPV-based materials is the diversity of organic molecules that can be used to design building blocks of donors, acceptors, and their interfaces.
Unfortunately, the experimental search for optimal OPV components is costly and very time-consuming. 
Thus, reliable quantum chemical methods combined with an efficient and flexible software package design are of utmost importance in the search for new, more efficient building blocks of OPV materials.~\cite{risko2011, cui-osc-review-2020}  
Commonly available models, like Density Functional Approximations (DFAs~\cite{yang_parr_book}), do not always provide reliable results for large extended $\pi$-systems,~\cite{pi-system-dft-failure} common building blocks of OPVs,~\cite{pi-extended-for-opv} and lack systematic improvability.~\cite{risko2011}
Difficulties originate from their electronic structures' biradical or multi-reference nature, which is often remarkably challenging to describe within a single-reference framework.  
Geminal-based methods are a promising alternative to conventional quantum chemistry models, providing a more compact representation of the correlated wave function.~\cite{surjan_1999, surjan2012, pawel-pccp-geminal-review-2022, johnson_2013, johnson2017strategies} 
Commonly used examples are the wave function classes based on the Antisymmetric Product of 1-reference orbital Geminal (AP1roG~\cite{limacher_2013, oo-ap1rog}), also known as pair Coupled Cluster Doubles (pCCD~\cite{tamar-pccd}), the Antisymmetrized Product of Strongly orthogonal Geminals (APSG~\cite{hurley_1953}), the Generalized Valence Bond (GVB)~\cite{gvb-pp-1}, and their orbital optimized variants.~\cite{oo-apsg, oo-ap1rog, piotrus_mol-phys, ap1rog-jctc, ps2-ap1rog} 
Combined with a reliable a posteriori correction to account for the missing dynamic correlation effects~\cite{pernal2014,apsg-erpa-pccp-2015, apsg-erpa-jctc-2014,pccd-ptx, ap1rog-lcc}, they allow us to model electron correlation effects effectively and in a balanced way.~\cite{post-pccd-entanglement} 
Yet, some further methodological developments are indispensable to meet the requirements of OPV applications and their transferability to other organic-based electronic molecules like those encountered in DSSCs.
Below, we list the key challenges that geminal-based methods have to overcome to become a potential driving force in the design of modern organic electronic building blocks and demonstrate some initial numerical examples.
%%%%%%%%%%%%%%%%%%%%%%%%%%%%%%%%%%%%%%%%%%%%%%%%%%%%%%%%%%
%% orbital energies
%%%%%%%%%%%%%%%%%%%%%%%%%%%%%%%%%%%%%%%%%%%%%%%%%%%%%%%%%%
\\ \textbf{Orbital energies}. 
OPVs can, in general, generate electricity from sunlight if the energy of light is equal to or greater than the donor-acceptor band gap offset. 
Thus, a critical factor in designing novel organic-based donor and acceptor molecules is the knowledge of the energies of the Highest Occupied Molecular Orbital (HOMO) and the Lowest Unoccupied Molecular Orbital (LUMO) and the corresponding HOMO--LUMO gap.
While these are easily obtained from the Hartree--Fock or DFA-based methods employing Koopman's or Janak's theorems, respectively,  they are not always reliable. 
Most geminal theories work with natural orbitals, where a (natural) occupation number is associated with each orbital, and orbital energies are not directly available. 
They must be deduced from the existing wave function, simultaneously ensuring efficiency and accuracy.
One of the simplest approximations uses information about the diagonal elements of the Fock matrix and the electron repulsion energy.~\cite{piotrus-orbital-energies}
More reliable orbital energies can be obtained from the Ionization Potential (IP)~\cite{Nooijen1992-ip, Nooijen1993-ip, Stanton1994-ip, Stanton1999-ip} and Electron Affinity (EA)~\cite{deaeomccsdt} variants of Equation-of-Motion (EOM)~\cite{rowe-eom, bartlett-eom, eom-cc-bartlett2012} applied on top of a geminal reference wave function~\cite{ip-pccd}.
From the IPs and EAs, we can deduce what is called the charge gap in the solid-state community (also referred to as the fundamental gap or HOMO--LUMO gap),
\begin{equation}\label{eq:charge-gap}
    \Delta_{\textrm c} = \textrm{IP} - \textrm{EA}.
\end{equation}
However, recent benchmark studies using the IP-EOM-pCCD approach~\cite{ip-eom-pccd-benchmark-pccp-2023} unravel the importance of dynamic correlation energy in reproducing reference data for small, compact organic molecules.
Thus, we require further methodological development and steeper computational scaling. 
The remedy seems to be deriving and calculating orbital energies from the extended Koopman’s theorem (EKT) on top of orbital-optimized methods.~\cite{ekt-day, ekt-levy, bozkaya2013-ekt} 
The approximate orbital energies are then determined by solving the secular equation 
\begin{equation}\label{eq:secular-ekt}
{\bf FC} -\epsilon {\bf \gamma C}  ={\bf 0}, 
\end{equation}
where ${\bf C}$ is the matrix of eigenvectors, $\epsilon$ is the diagonal matrix of eigenvalues (orbital energies), $\gamma_{pq}$ is the one-particle reduced density matrix (1-RDM), 
\begin{equation}\label{eq:1-rdm}
\gamma_{pq}=\langle \Psi^N|\hat{a}_p^{\dagger}\hat{a}_q|\Psi^N \rangle, 
\end{equation}
and ${\bf F}$ is the so-called generalized Fock matrix (aka Lagrangian), 
\begin{equation}
\text{F}_{pq}=-\langle \Psi^N|\hat{a}_p^{\dagger}[\hat{H},\hat{a}_q]|\Psi^N\rangle
=\sum_r h_{pr} \gamma_{qr} + 2\sum_{r,s,t} g_{prst}  \Gamma_{qrst}, 
\end{equation}
which can be expressed in terms of one- and two-electron integrals, $h_{pr}$ and $g_{prst}$, and reduced density matrices. 
In the above expression, the two-particle reduced density matrix (2-RDM) is defined as
\begin{equation}
\Gamma_{pqrs} = \langle \Psi^N|\hat{a}_p^{\dagger}\hat{a}_q^{\dagger} \hat{a}_s \hat{a}_r |\Psi^N \rangle.
\end{equation}
Thus, to calculate (approximate) orbital energies from orbital-optimized geminal-based methods, the corresponding 1- and 2-RDMs must be reliable. 
An alternative formulation of EKT, proposed by Ciosłowski and coworkers,~\cite{cioslowski1997-ekt} uses energy-derivative density matrices.  
%%%%%%%%%%%%%%%%%%%%%%%%%%%%%%%%%%%%%%%%%%%%%%%%%%%%%%%%%%
%% properties
%%%%%%%%%%%%%%%%%%%%%%%%%%%%%%%%%%%%%%%%%%%%%%%%%%%%%%%%%%
\\ \textbf{Electronic properties}. 
The power conversion efficiency of OPV devices can be improved by exploiting building blocks that feature a complementary spectral absorption range between the donor and the acceptor and a strong absorption in the visible-near infrared region to ensure a large short-circuit current. 
To predict such properties through large-scale quantum chemical modeling, we must efficiently and reliably determine the electronic spectra (electronic excitation energies and associated transition dipole moments) and ground-state electronic properties like electronic dipole and quadrupole moments.
Optimizing these properties is indispensable for improving charge separation, transport, and overall device performance in organic solar cells.
Having computed the lowest excitation energy, also denoted as the optical gap ($\Delta_{\textrm o}$), and the charge gap (eq.~\eqref{eq:charge-gap}), we can determine the exciton binding energy~\cite{ebe_perspective} 
\begin{equation}
    \textrm{EBE} = \Delta_{\textrm c} - \Delta_{\textrm o}. 
\end{equation}
The EBE denotes the energy required to dissociate an excited electron-hole pair into free charge carriers. 
Specifically, the exciton is formed in the donor domain of the OSC. At the same time, the acceptor material is meant to provide a way to overcome the corresponding EBE and, hence, separate the charges.
Thus, we should be able to predict reliable EBEs for the donor and the donor--acceptor interface to steer the efficiency of OPV devices.

\begin{figure}
    \centering
    \caption{The molecular structures of (a) naphthalene diimide (NDI), two carbazole-based dyes with different $\pi$-bridge units such as (b) benzodithiophene (CBA) and (c) dithienopyrrole (CDA). For structures (b) and (c), the donor, $\pi$-bridge, and acceptor components are highlighted with blue, green, and red rectangles, respectively. Structure (a) was relaxed using B3LYP/cc-pVDZ, while (b) and (c) were optimized using M06/6-31G(d), and their geometries are available in the SI of Ref.~\citenum{delgado-lena-reference}.}
    \label{fig:structures}
    \includegraphics[width=1.0\columnwidth]{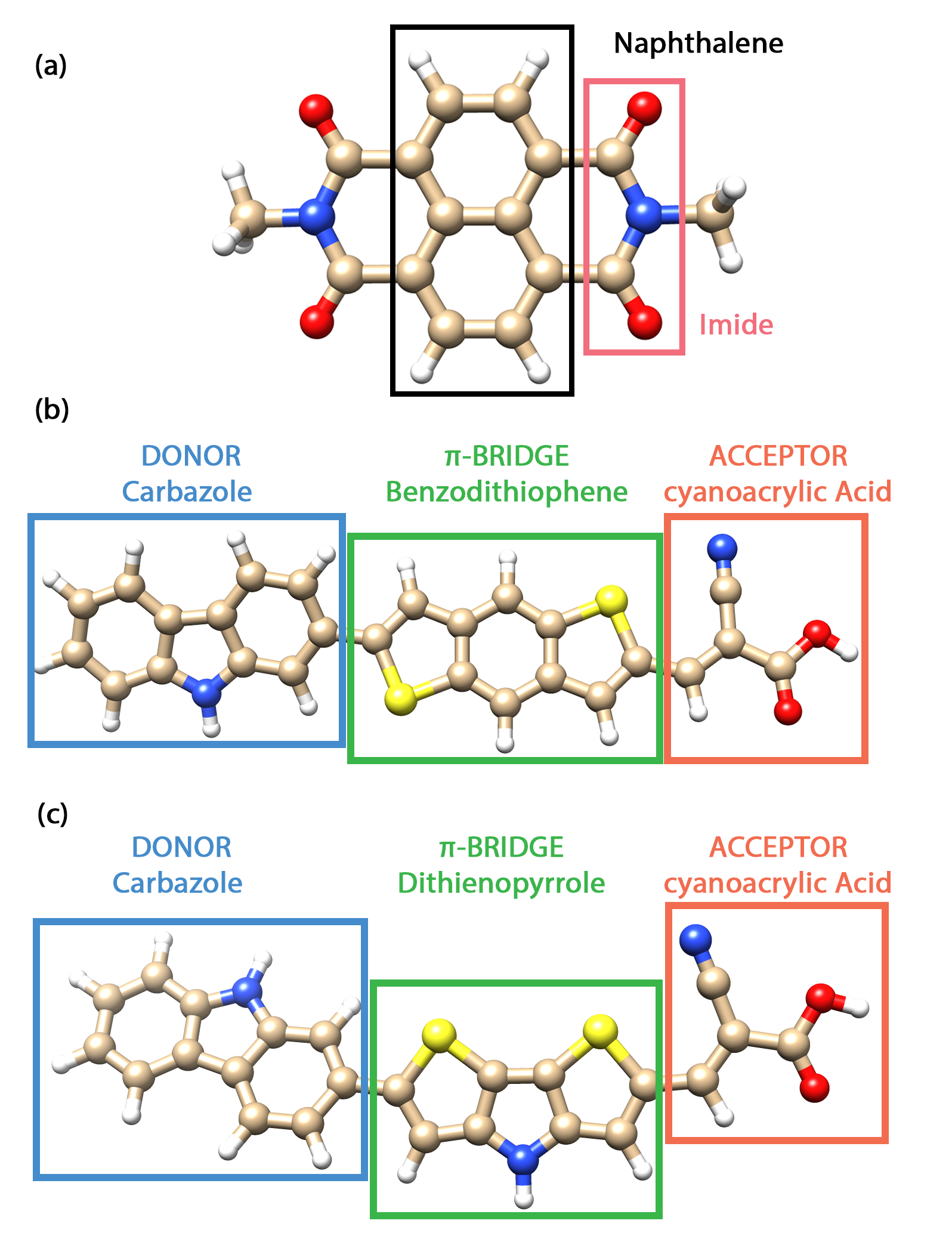}
\end{figure}

Most promising geminal theories for excited state calculations are based on the extended random phase approximation~\cite{gvb-erpa, apsg-erpa-pccp-2015} and the EOM formalism.~\cite{eom_pccd, eom_pccd_erratum, eom-pccd-lccsd} 
Although most EOM-based methods yield size-intensive energies, the corresponding properties derived from transition density matrices are not. 
Yet, the computations of transition dipole moments from EOM could be more computationally impractical due to the need to compute both left and right eigenvectors. 
A remedy to this problem is linear-response theory, which can be used in large-scale modeling.~\cite{hapka-jensen}
Finally, we should stress that a reliable prediction of electronic dipole and quadrupole moments from geminal theories requires the inclusion of single excitations in the theoretical model. 
%%%%%%%%%%%%%%%%%%%%%%%%%%%%%%%%%%%%%%%%%%%%%%%%%%%%%%%%%%
%% dispersion
%%%%%%%%%%%%%%%%%%%%%%%%%%%%%%%%%%%%%%%%%%%%%%%%%%%%%%%%%%
\\ \textbf{Non-covalent interactions}.
%%%%%%%%%%%%%%%%%%%%%%%%%%%%%%%%%%%%%%%%%%%%%%%%%%%%%%%%%%
The materials of OPV's active layer should exhibit suitable aggregation properties to form nanoscale phase separations and interpenetrating networks. 
Thus, quantum chemistry methods must accurately predict the intermolecular interaction energies between large molecules.
Such a task is remarkably difficult because the interaction energy often features a considerable amount of non-covalent/dispersion interactions, which are challenging to model reliably employing methods designed for strong electron correlation. 
An exception are geminal-based approaches, which proved to be reliable and computationally efficient in describing systems featuring a mixture of non-dynamic and non-covalent interactions.~\cite{surjan-bond-1985, hapka-jctc-2018,hapka-jctc-2021,hapka-jensen, filip-jctc-2019, dispersion-visualization-jpca-2022} 
Apart from Symmetry Adapted Perturbation Theory~\cite{sapt-chem-rev-1994} and a linearized coupled-cluster correction on top of a geminal reference function~\cite{zoboki2013, filip-jctc-2019}, new geminal-based models are highly desirable to facilitate modeling of non-covalent interactions and large molecules. 
A promising approach to theoretically describe large-scale non-covalent interactions in OPVs is a hybrid method that combines a given geminal ansatz with a semi-classical dispersion correction.
In such models, the geminal part captures long-range electron correlation effects, while short-range dynamic electron correlation is handled by DFAs and a semi-classical dispersion correction to account for long-range dynamic correlations.~\cite{garza-pccp-wdv} 
Furthermore, the commonly used exchange--correlation functionals can be combined with a D3 dispersion correction and Becke--Johnson dumping.~\cite{grimme2011-d3}
%%%%%%%%%%%%%%%%%%%%%%%%%%%%%%%%%%%%%%%%%%%%%%%%%%%%%%%%%%
%% embedding and model Hamiltonians
%%%%%%%%%%%%%%%%%%%%%%%%%%%%%%%%%%%%%%%%%%%%%%%%%%%%%%%%%%
\\ \textbf{Quantum embedding with geminals}. 
Another challenge the quantum chemical modeling of realistic OPVs faces is the need to cope with many atoms.
A promising approach to circumvent this problem and reduce the computational cost dramatically originates from quantum embedding techniques.~\cite{gomes_rev_2012} 
Its idea relies on the fact that electron correlation is \textit{`local'} in nature,~\cite{pulay1983localizability, hampel1996local} allowing us to partition the whole system into subsets.~\cite{pernal-rdm-emb-pccp-2016, gvb-embedding-jctc-2017}
Examples are the WFT-in-DFT and WFT-in-WFT approaches, where a geminal wave function treats only a small subset of atoms in the whole molecular structure. In contrast, the remaining part is treated with a computationally more efficient, albeit less accurate, method.
Such approaches effectively account for environmental effects also present in OPVs. 
So far, only the simplest static embedding model has been tested for geminals.~\cite{pccd-static-embedding}
More reliable embedding schemes exploiting orbital optimization procedures within geminals, similar to the \textit{`freeze-and-thaw'} protocol,~\cite{fde-freez-and-thaw} are yet to be developed. 
%%%%%%%%%%%%%%%%%%%%%%%%%%%%%%%%%%%%%%%%%%%%%%%%%%%%%%%%%%
%% entanglement
%%%%%%%%%%%%%%%%%%%%%%%%%%%%%%%%%%%%%%%%%%%%%%%%%%%%%%%%%%
\\ \textbf{Quantum entanglement and correlation}. 
Due to the restriction to electron-pair excitations, the corresponding 1- and 2-RDMs can be calculated computationally more effectively than in conventional wave function-based methods. 
The approximate 1- and 2-RDMs obtained from (orbital-optimized) geminal-based wave functions can be used to calculate the single- and two-orbital entropies from which quantum entanglement and correlation measures can be computed.~\cite{rissler2006, entanglement_letter, entanglement-jctc-2013, kasia-orbital-entanglement-ijqc-2015, ijqc-eratum, Ding2020, qit-concepts-schilling-jctc-2021}
The single-orbital entropy determines the quantum entanglement between each orbital and the orbital bath and can be applied to deduce molecular bond-orders~\cite{entanglement-jctc-2013, pccp_bonding, pawel-pccp-2015} and guide the partitioning of the quantum system.~\cite{cuo_dmrg, kasia-orbital-entanglement-ijqc-2015, boguslawski2017, stein2016, pccd-static-embedding} 
The so-called mutual information allows us to dissect electron correlation into different types~\cite{entanglement_letter} and quantify the interaction between orbitals in the system (in terms of orbital-pair correlations).~\cite{cuo_dmrg, ola-qit-actinides-pccp-2022}  
Altogether, these tools provide a deeper understanding of electronic structures using the language of interacting orbitals.
Being able to scrutinize the interactions of the HOMO and LUMO orbitals with the remaining molecular orbitals might guide the optimization process of developing new organic solar cells and their building blocks.
%%%%%%%%%%%%%%%%%%%%%%%%%%%%%%%%%%%%%%%%%%%%%%%%%%%%%%%%%%
%% interpretational power
%%%%%%%%%%%%%%%%%%%%%%%%%%%%%%%%%%%%%%%%%%%%%%%%%%%%%%%%%%
\\ \textbf{Interpretational potential}.
Since geminal-based methods exploit two-electron functions as the fundamental building blocks of the electronic wave function, the corresponding molecular orbital basis (used to construct each geminal) is typically localized on just a few centers of the whole molecule.
%That is, geminal-based approaches entail interpreting computational results within a localized molecular orbital basis.
Due to these localized molecular orbitals, the contributions to excited, ionized, or electron-attached states feature several contributions (Slater determinants) with small weights.
Thus, the underlying electronic structure differs from the conventional picture we obtain when working with delocalized canonical orbitals like those predicted by DFAs.
The significant advantage of a localized basis~\cite{stewart-interpretation-of-localized-orbitals-2019, local-orbitals-in-quantum-chemistry} is the clear distinction of the donor and acceptor regions or their interfaces.~\cite{delaram-pani-pccd-2023} Electronic excitations can be unambiguously assigned to specific molecular basins allowing to dissect the electronic excitations into, for instance, charge transfer or local ones, while HOMO and LUMO orbitals can be located on, for instance, donor or acceptor domains.
Such an analysis is particularly beneficial in designing OPV building blocks that are desired to feature the HOMO/LUMO on specified domains or increase the charge-transfer character in the excited states of interest.
%%%%%%%%%%%%%%%%%%%%%%%%%%%%%%%%%%%%%%%%%%%%%%%%%%%%%%%%%%
%% software
%%%%%%%%%%%%%%%%%%%%%%%%%%%%%%%%%%%%%%%%%%%%%%%%%%%%%%%%%%
\\ \textbf{Interoperable and reusable software}.
A key factor in developing new methods, like geminals, is their implementation in modern software packages, fulfilling the desired FAIR (Findable, Accessible, Interoperable, and Reusable) features.~\cite{fair}
While many geminal methods are scattered across different software platforms and packages,~\cite{pybest-paper, ghent-qc, hapka-jensen} the interoperability and reusability of geminal-based software are indispensable for their faster development, testing, and application to OPV-related problems.
Geminal-based methods could become a driving force for quantum chemical modeling of organic electronic compounds or their building blocks if such requirements are genuinely fulfilled.  
%%%%%%%%%%%%%%%%%%%%%%%%%%%%%%%%%%%%%%%%%%%%%%%%%%%%%%%%%%
%% numerical exampls
%%%%%%%%%%%%%%%%%%%%%%%%%%%%%%%%%%%%%%%%%%%%%%%%%%%%%%%%%%
\\ \textbf{Numerical examples}.
In the following, we illustrate the performance of orbital-optimized pCCD methods in predicting various molecular properties in three organic molecules depicted in Fig. \ref{fig:structures}.
These molecules represent examples of how to advance the performance of organic electronic components:
Naphthalene diimide (NDI) as an excellent acceptor (see Fig~\ref{fig:structures}(a)) and two carbazole-based dyes of DSSCs proposed in Ref.~\citenum{delgado-lena-reference}, namely CBA (see Fig~\ref{fig:structures}(b)) and CDA (see Fig~\ref{fig:structures}(c)), respectively.
The electronic properties of those molecules in terms of the HOMO and LUMO energies, IP and EA, and $\Delta_{\textrm c}$ calculated from the IP/EA-EOM-pCCD models using the {1h/1p} (1 hole/1 particle) and {2h1p/2p1h} (2 holes 1 particle/2 particles 1 hole) operators are presented in Table~\ref{tab:energies}.
For NDI, we compare the pCCD results to the Restricted Hartree-Fock (RHF) and B3LYP~\cite{b3lyp}, and experimental results.
For the carbazole-based dyes, the performance of pCCD is compared to PBE~\cite{pbe01, pbe02} (GGA functional), PBE0~\cite{pbe0} (hybrid functional with 25\% of HF exchange), CAM-B3LYP\cite{cam-b3lyp} (range-separated hybrid exchange--correlation functional with 19\% and 65\% of HF exchange for the short and long-range, respectively), and the statistical average of orbital model exchange--correlation potential (SAOP).~\cite{saop}

%%%%%%%%%%%%%%%%%%%%%%%%%%%%%%%%%%%%%%%%%%%%%%%%%%%%%%%%%%
%% NDI
%%%%%%%%%%%%%%%%%%%%%%%%%%%%%%%%%%%%%%%%%%%%%%%%%%%%%%%%%%
NDI, a member of the rylene diimide molecule class,~\cite{rylene-adv-mat-2011,perylene-advance-materials-2012,pyrelene-rd1-reference,perylene-chem-com-2018} is composed of a naphthalene core---a conjugated $\pi$ system---linked at both termini to imide units (Fig.~\ref{fig:structures}(a)).
NDI and its derivatives showcase excellent n-type semiconductor characteristics, rendering them suitable acceptors in organic solar cells (OSCs) due to their robust electron acceptor properties, high chemical and thermal stability, substantial absorption coefficient, and prominent fluorescence.~\cite{nbi-review}
Theoretical investigations~\cite{rylenediimides} demonstrated that the B3LYP functional yields HOMO and LUMO energies in good alignment with experimental values. The overestimation of the HOMO energy by 0.47 eV leads to a larger HOMO--LUMO gap of 0.49 eV compared to experimental findings.
Significantly poorer results emerged with the RHF method, which decreases the HOMO energy by 1.96 eV and increases the LUMO energy by 3.2 eV relative to B3LYP results.
This shift produces a massive $\Delta_{\textrm c}$ energy of 8.80 eV.
IP/EA-EOM-pCCD(1h/1p) yields even more significant HOMO and LUMO energy errors.
If the IP/EA description is treated more accurately and extended to the 2h1p/2p1h level, the HOMO energy aligns excellently with experimental results (within chemical accuracy), while the LUMO is underestimated by 0.52 eV.
The larger error of the LUMO level increases the $\Delta_{\textrm c}$ gap by 0.59 eV compared to experiment.
Note that the EA-EOM formalism is more sensitive to the basis set size and environmental effects;~\cite{eom-exciton-binding-energy-faraday-diss-2014} hence, larger errors in LUMO energies are expected.

\begin{table}
    \caption{Performance of various quantum chemistry methods in predicting HOMO, LUMO, and $\Delta_{\textrm c}$ energies (in eV). }
    \begin{tabular}{l r r r }
    \hline
                     & HOMO & LUMO & $\Delta_{\textrm c}$ \\
    \hline
    \textbf{NDI}\\
    B3LYP/6-311g(d, p)\cite{rylenediimides}          & $-$7.25     & $-$3.61     &  3.64       \\  
    RHF/cc-pVDZ                                      & $-$9.21     & $-$0.41     &  8.80       \\ 
    IP/EA-EOM-pCCD(1h/1p)/cc-pVDZ                    & $-$9.89     &  0.42       &  10.31      \\ 
    IP/EA-EOM-pCCD(2h1p/2p1h)/cc-pVDZ                & $-$6.85     & $-$3.11     &  3.74       \\ 
    Experimental\cite{rylenediimides}                & $-$6.78     & $-$3.63     &  3.15       \\
    & &  &   \\
    \textbf{CBA}   \\
    PBE/TZ2P                                          & $-$5.46     & $-$3.72     &   1.74      \\
    PBE0/TZ2P                                         & $-$6.23     & $-$3.07     &   3.16      \\
    CAM-B3LYP/TZ2P                                    & $-$7.20     & $-$2.26     &   4.94      \\
    SAOP/TZ2P                                         & $-$9.51     & $-$1.69     &   7.82      \\ 
    IP/EA-EOM-pCCD(1h/1p)/cc-pVDZ                     & $-$7.86     &  0.90       &   8.76      \\ 
    IP/EA-EOM-pCCD(2h1p/2p1h)/cc-pVDZ                 & $-$4.91     & $-$2.59     &   2.32      \\ 
    & &  &   \\
   \textbf{CDA}   \\ 
   PBE/TZ2P                                           & $-$5.27     & $-$3.47     &   1.80      \\
   PBE0/TZ2P                                          & $-$6.00     & $-$2.94     &   3.06      \\
   CAM-B3LYP/TZ2P                                     & $-$6.94     & $-$2.06     &   4.88      \\
   SAOP/TZ2P                                          & $-$9.23     & $-$7.45     &   1.78      \\ 
   IP/EA-EOM-pCCD(1h/1p)/cc-pVDZ                      & $-$7.51     &  1.05       &   8.56      \\ 
   IP/EA-EOM-pCCD(2h1p/2p1h)/cc-pVDZ                  & $-$4.51     & $-$2.40     &   2.11      \\ 
    \hline
    \end{tabular}
    %\caption{}
    \label{tab:energies}
\end{table}

%%%%%%%%%%%%%%%%%%%%%%%%%%%%%%%%%%%%%%%%%%%%%%%%%%%%%%%%%%
%% CBA and CDA
%%%%%%%%%%%%%%%%%%%%%%%%%%%%%%%%%%%%%%%%%%%%%%%%%%%%%%%%%%

\begin{figure}
    \centering
    \includegraphics[width=\columnwidth]{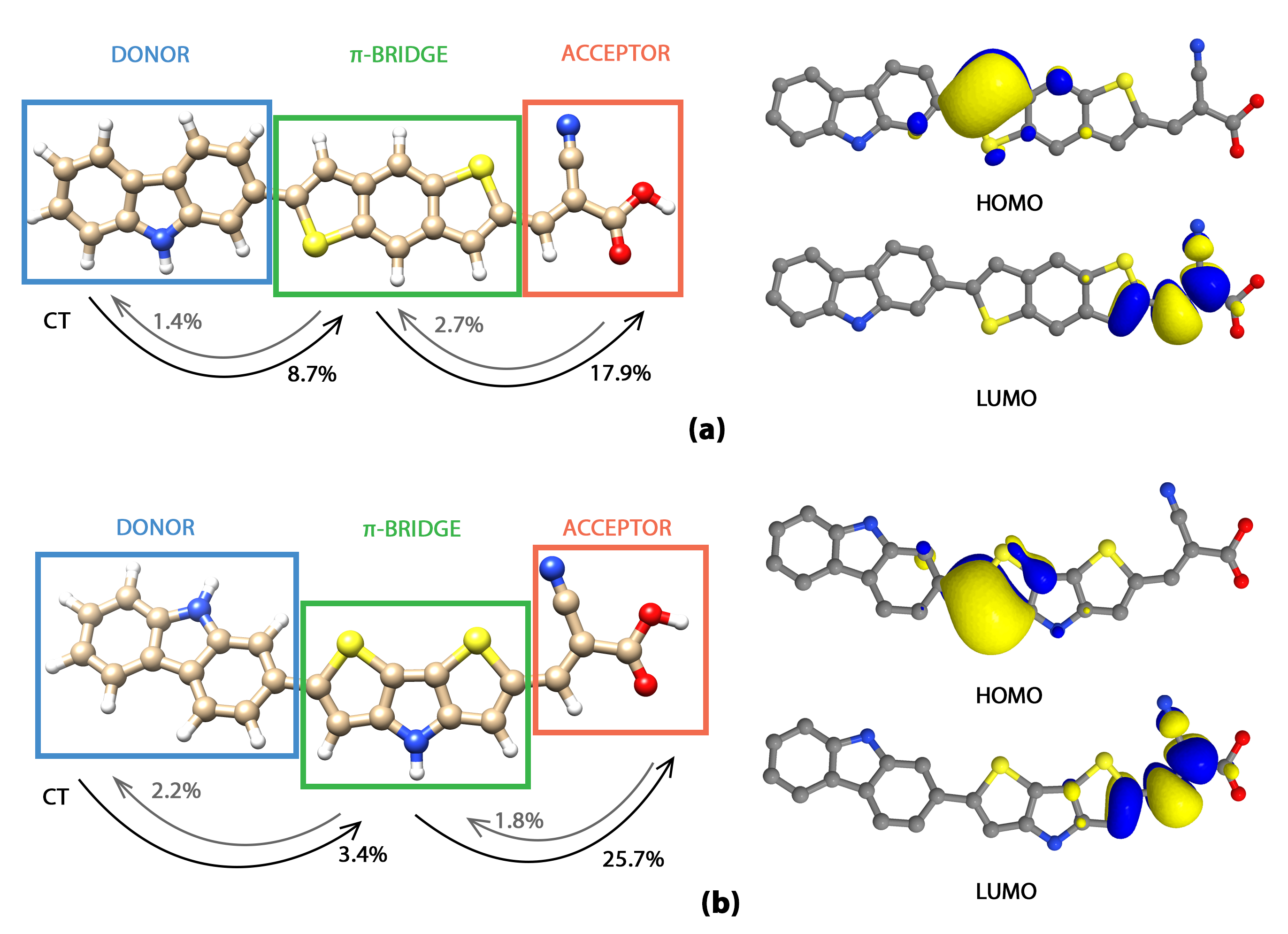}
    \caption{The pCCD HOMO and LUMO orbitals of the (a) CBA and (b) CDA dyes. On the left-hand side, the percentage of charge transfer character to the lowest-lying excited state for each dye is included.}
    \label{fig:dyes}
\end{figure}

The two carbozyle-based dyes shown in Fig.~\ref{fig:structures}(b) and (c) were proposed in Ref.~\citenum{delgado-lena-reference} as new donor--$\pi$-bridge--acceptor organic sensitizers for DSSCs.
These dyes incorporate a carbazole donor and a cyanoacrylic acid acceptor while utilizing benzodithiophene and dithienopyrrole as the $\pi$-spacers in CBA and CDA, respectively.
Accurately determining the energy alignment between a sensitizer and a semiconductor substrate (such as TiO$_2$) is pivotal in assessing the applicability of a dye in DSSCs.
Hence, the precise determination of the HOMO and LUMO energies is paramount.
As indicated in Table~\ref{tab:energies}, the HOMO, LUMO, and $\Delta_{\textrm c}$ energies strongly rely on the fraction of HF exchange and the choice of DFA.
For instance, for the CBA dye, the HOMO and LUMO energies vary by up to 4.05 and 2.03 eV, respectively.
Such substantial discrepancies in the frontier orbital energies notably affect $\Delta_{\textrm c}$, ranging from 1.74 eV to 7.82 eV.
Like NDI, $\Delta_{\textrm c}$ determined by IP/EA-EOM-pCCD(1h/1p) is significantly reduced when the 2h1p/2p1h sectors are employed in the IP/EA models.
Finally, we should note that the localized nature of the pCCD-optimized orbitals allows us to locate the HOMO and LUMO across the dye sensitizer:
While the LUMO of both CBA and CDA is located on the acceptor domain, the HOMO is centered mainly on the $\pi$ bridge (see Fig.~\ref{fig:dyes}).

In contrast to the frontier orbital energies, the discrepancies in the lowest-lying excited state energies are much less pronounced if DFAs are concerned, differing by a maximum of 1.07 eV across the investigated functionals (2.08/2.32 for PBE, 2.66/2.75 for PBE0, 3.11/2.88 for CAM-B3LYP, and 2.04/2.23 for SAOP for CBA/CDA, respectively).
The first excitation energy further increases in EOM-pCCD+S calculations to 4.76 eV for CBA and 4.36 eV for CDA, respectively.
A detailed analysis of the excited state wave function at the EOM-pCCD+S level of theory, considering about 90\% of the configurational weights, suggests that the lowest-lying excited state of the CBA dye features around 30\% charge transfer character, involving electronic transitions between donor and acceptor units, from which the majority (87\%) entails charge transfer from the donor to the acceptor unit (see also Fig.~\ref{fig:dyes}(a)).
In contrast, the charge transfer character in the CDA molecule increases to 33\%, with a similar portion of 88\% going from the donor to the acceptor domain (see also Fig.~\ref{fig:dyes}(b)).
A similar analysis, considering only 71\% (CBA) or 74\% (CDA) of the configurational space, is included in the SI.

Thus, our numerical examples illustrate that pCCD-based methods can predict accurate molecular properties (HOMO/LUMO energies and charge/band/fundamental gaps). In contrast, the localized nature of the molecular orbital basis allows us to unambiguously identify the domain on which the HOMO or LUMO orbitals are centered.
Furthermore, each excited state wave function can be broken down into electronic excitations centered on specific domains, like local or CT excitations.
Although such an analysis might initially seem tedious, it is possible to fully automatize the assessment of the excitation characters by assigning molecular orbitals to specific domains.
Finally, another challenge of geminal-based methods will be pushing the precision to chemical accuracy, considering environmental effects or dynamical correlation.

%%%%%%%%%%%%%%%%%%%%%%%%%%%%%%%%%%%%%%%%%%%%%%%%%%%%%%%%%%
%% Acknowledgement
%%%%%%%%%%%%%%%%%%%%%%%%%%%%%%%%%%%%%%%%%%%%%%%%%%%%%%%%%%
\section{Acknowledgement}
P.T.~acknowledges financial support from the SONATA BIS research grant from the National Science Centre, Poland (Grant No. 2021/42/E/ST4/00302) and the scholarship for outstanding young scientists from the Ministry of Education and Science. 
M.G.~thanks financial support from a Ulam NAWA -- Seal of Excellence research grant (no.~BPN/SEL/2021/1/00005). 
The research leading to these results has received funding from the Norway Grants 2014--2021 via the National Centre for Research and Development.

%%%%%%%%%%%%%%%%%%%%%%%%%%%%%%%%%%%%%%%%%%%%%%%%%%%%%%%%%%%%%%%%%%%%%
%% The same is true for Supporting Information, which should use the
%% suppinfo environment.
%%%%%%%%%%%%%%%%%%%%%%%%%%%%%%%%%%%%%%%%%%%%%%%%%%%%%%%%%%%%%%%%%%%%%
%\begin{suppinfo}

%This will usually read something like: ``Experimental procedures and
%characterization data for all new compounds. The class will
%automatically add a sentence pointing to the information on-line:

%\end{suppinfo}

%%%%%%%%%%%%%%%%%%%%%%%%%%%%%%%%%%%%%%%%%%%%%%%%%%%%%%%%%%%%%%%%%%%%%
%%%%%%%%%%%%%%%%%%%%%%%%%%%%%%%%%%%%%%%%%%%%%%%%%%%%%%%%%%%%%%%%%%%%%
\bibliography{rsc} %You need to replace "rsc" on this line with the name of your .bib file
\bibliographystyle{rsc} %the RSC's .bst file

\end{document}